\documentclass[twocolumn, tighten]{aastex63}
\usepackage{amssymb,esvect,amsmath,graphicx,latexsym,amsthm,slashed,eso-pic,cases}

\newcommand{\ep}{\epsilon}

\newcommand{\pL}{\left(} \newcommand{\pR}{\right)} \newcommand{\bL}{\left[} \newcommand{\bR}{\right]}    
\newcommand{\beq}{\begin{equation}} \newcommand{\eeq}{\end{equation}}
\newcommand{\bea}{\begin{eqnarray}} \newcommand{\eea}{\end{eqnarray}}

\newcommand{\Eq}[1]{Eq.~(\ref{#1})}  
  
\newcommand{\Fig}[1]{Fig.~\ref{#1}}

\newcommand{\msun}{{\rm M}_\odot}

\newcommand{\nbh}{N_{\rm BH}}
\newcommand{\mbh}{M_{\rm BH}}
\newcommand{\mbhmg}{M_{\rm BHMG}}
\newcommand{\mmin}{M_{\rm min}}

\newcommand{\mzams}{M_{\rm ZAMS}}
\newcommand{\mhb}{M_{\rm hb}}

\DeclareRobustCommand{\okina}{%
  \raisebox{\dimexpr\fontcharht\font`A-\height}{%
    \scalebox{0.8}{`}%
  }%
}

\begin{document}

\title{Find the Gap: Black Hole Population Analysis with an Astrophysically Motivated Mass Function}
\author{Eric J.~Baxter}
\affiliation{Institute for Astronomy, University of Hawai\okina i, 2680 Woodlawn Drive, Honolulu, HI 96822, USA}
\author{Djuna Croon}
\affiliation{TRIUMF, 4004 Wesbrook Mall, Vancouver, BC V6T 2A3, Canada}
\affiliation{Institute for Particle Physics Phenomenology, Department of Physics, Durham University, Durham DH1 3LE, U.K.}
\author{Samuel D.~McDermott}
\affiliation{Fermi National Accelerator Laboratory, Batavia, IL, USA}
\author{Jeremy Sakstein}
\affiliation{Department of Physics \& Astronomy, University of Hawai\okina i, Watanabe Hall, 2505 Correa Road, Honolulu, HI, 96822, USA}

\begin{abstract}
    We introduce a novel black hole mass function which realistically models the physics of pair instability supernovae with a minimal number of parameters. Applying this to all events in the LIGO-Virgo GWTC-2 catalog, we detect a peak at 
    $\mbhmg =46^{+17}_{-6} \,\msun$. 
    Repeating the analysis without the black holes from the event GW190521, we find this feature at 
    $\mbhmg = 54\pm 6\, \msun$.
    These results establish the edge of the anticipated ``black hole mass gap'' at a value compatible with the expectation from standard stellar structure theory. 
    The mass gap manifests itself as a discontinuity in the mass function and is populated by a distinct, less abundant population of higher-mass black holes.
    We find that the primary black hole population scales with power-law index 
    $-1.95\pm0.51$ ($-1.97\pm0.44$)
    with (without) GW190521, consistent with models of star formation. Using Bayesian techniques, we establish that our mass function fits a new catalog of black hole masses approximately as well as pre-existing phenomenological mass functions.
    We also remark on the implications of these results for constraining or discovering new phenomena in nuclear and particle physics.
    \hfill \textsc{fermilab-pub-21-148-t, ippp/20/88}
\end{abstract}

\keywords{
gravitational waves
 --- stars:black holes ---  
 astroparticle physics
}

\section{Introduction}
With the release of GWTC-2 \citep{Abbott:2020niy}, the LIGO-Virgo collaboration (LVC) has enabled a dramatically new understanding of the contents of the cosmos. As the total number of binary merger detections increases, our understanding deepens, and the possibilities for learning more about the fundamental constituents of the Universe have vastly expanded. For example, the GW190521 event \citep{Abbott:2020tfl} established a qualitatively new range of masses for compact objects, enabling the study of intermediate mass black holes for the first time \citep{Abbott:2020mjq}. 

With the GWTC-2 dataset, population studies of black holes have become more informative \citep{Abbott:2020gyp}. 
Such aggregate studies can enable a new understanding of stellar dynamics \citep{Fishbach:2017zga}. In particular, the physics of pair-instability supernovae (PISN) is expected to introduce features in the distribution of black hole masses \citep{2016A&A...594A..97B, 2017ApJ...840L..24F, 2017PhRvD..95l4046G, Talbot:2018cva, Wang2020arXiv200903854W, Wang2021arXiv210402566W}.
Most importantly, PISN leave no compact remnant for a wide range of initial stellar masses: 
this unpopulated space in the stellar graveyard is known as the black hole mass gap (BHMG). 

In this work, we establish a minimal (three-parameter) black hole mass function of first-generation black holes that 
includes the signature of the astrophysical pair instability. This function has a single dimensionful parameter, $\mbhmg$, which reveals the location of the lower edge of the BHMG. Our mass function can therefore be used to directly extract physically meaningful constraints on the BHMG from data.  In contrast, several recent analyses have adopted phenomenological models for the BHMG that do not directly relate to the predictions of stellar structure theory. We apply our model to the GWTC-2 catalog, allowing for a subdominant two-parameter ``pollutant'' population with masses within the BHMG, and, following \citet{Abbott:2020gyp}, a two-parameter model of low-mass black hole ``formation efficiency.''
We establish that our black hole mass function fits the data approximately as well as the best mass function proposed in \citet{Abbott:2020gyp}, but with the added benefit of enabling a transparent interpretation of the single dimensionful degree of freedom in our fit.

\begin{figure*}[t]
\begin{center}
\includegraphics[width=.485\textwidth]{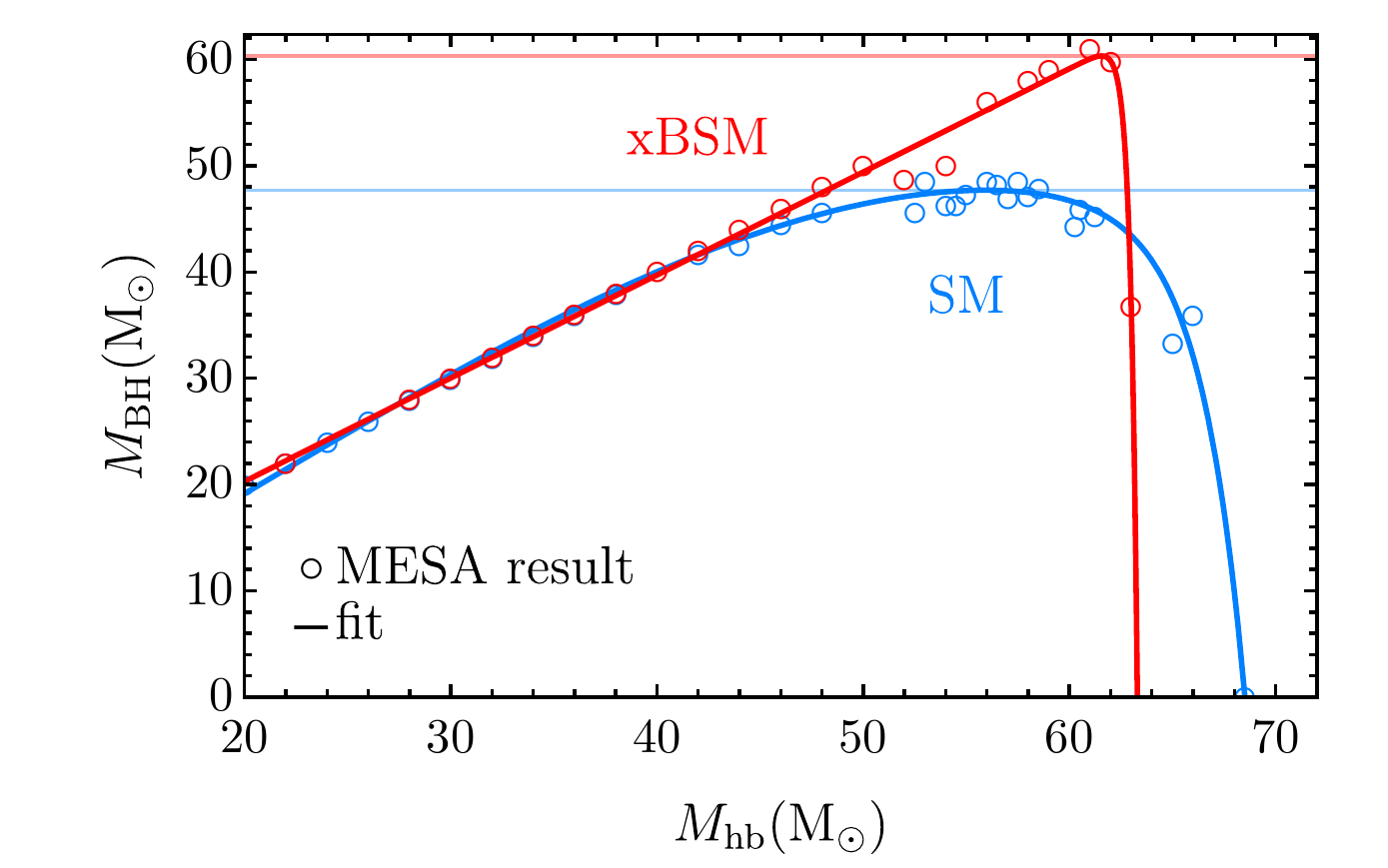}
\includegraphics[width=.485\textwidth]{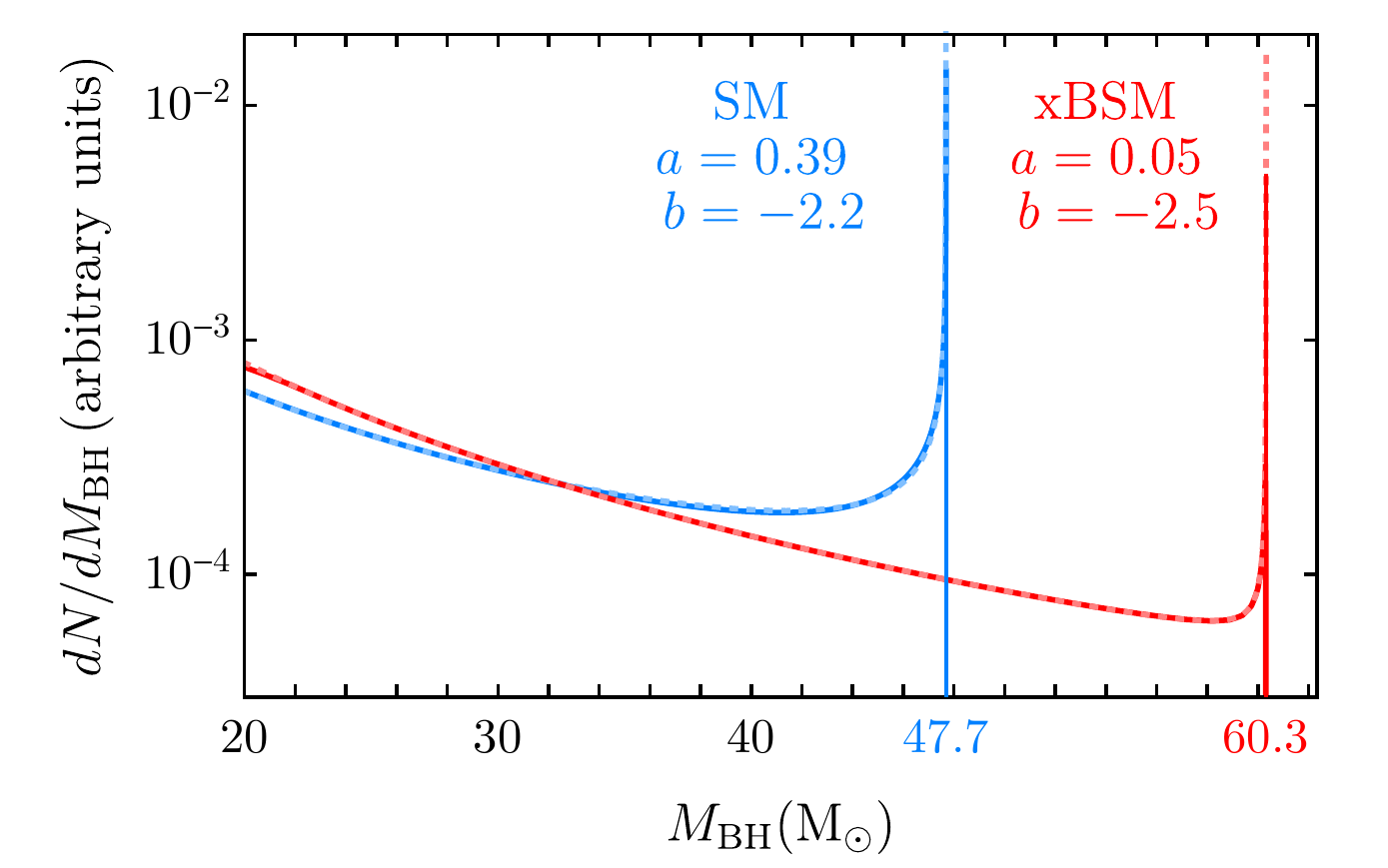}
\includegraphics[width=.485\textwidth]{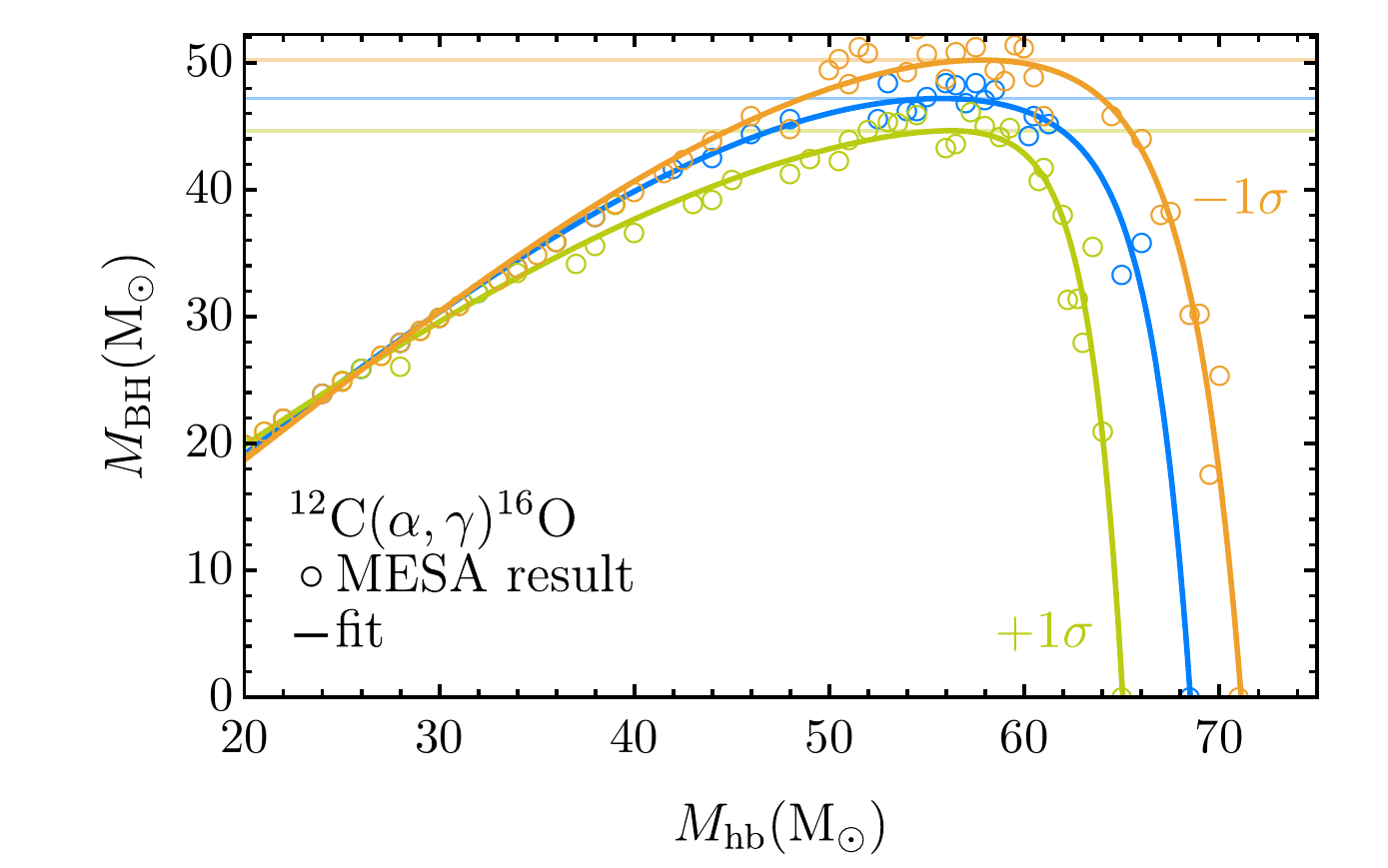}
\includegraphics[width=.485\textwidth]{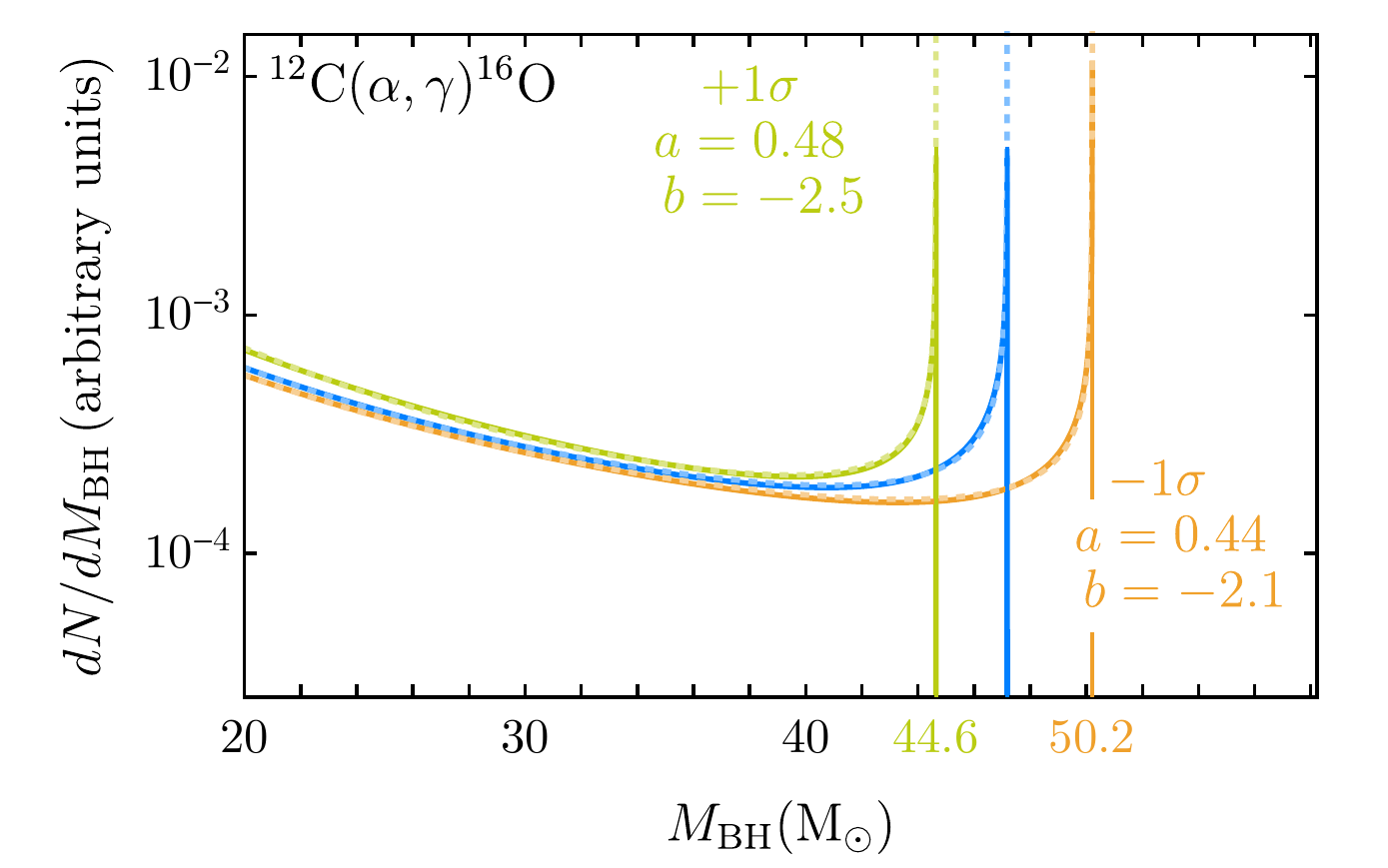}
\caption{{\it Left:} Stellar remnant masses as found with {\tt MESA} with best fits as described in the text. Here ``xBSM" is an example of a beyond the Standard Model scenario which avoids PPISN --- more detail can be found in the text and in \citet{Croon:2020oga}.
The rate for the ${}^{12}{\rm C} (\alpha, \gamma)^{16}{\rm O}$ reaction is implemented using the latest data from \cite{deBoer2017RvMP...89c5007D} (tabulated in the reproduction package of \cite{2020ApJ...902L..36F}; see App.~\ref{app:rateCO}). We simulate stars with $Z=10^{-5}$. {\it Right:} Resulting black hole mass functions. The dashed lines are the fit using Eq.~\eqref{nbh-1g} and best-fit parameters $a$ and $b$ are given in the figure. Here the IMF is assumed to be a power law with index $ -2.4$. 
}
\label{fig:mesaMF}
\end{center}
\end{figure*}

\section{Astrophysical mass function}
In complete generality, we can write the mass function of ``first-generation'' black holes arising from isolated stellar evolution as
\beq \label{chain-rule-1}
\frac{d\nbh^{\rm(1g)}}{d \mbh} = \int d \vec \theta \, \frac{dN_*}{dM_*}   \frac{dM_*(\vec \theta)}{d \mbh} \cal P(\vec \theta),
\eeq
where: $M_*$ is a stellar mass and $\vec \theta$ are nuisance parameters such as metallicity, redshift, binarity, or rotation 
\citep{Farmer:2019jed, 2019ApJ...882...36M, 2020A&A...640L..18M, Woosley:2021xba}; and $\cal P(\vec{\theta})$ is the probability of the parameters $\vec \theta$ across the stellar population. Because of pulsational pair instability supernovae (PPISN), the relation between $M_*$ and $\mbh$ is not bijective: pulsational mass loss triggered by the electron-positron pair instability means that stars of different initial stellar masses can form identical mass black holes. However, the function $\mbh(M_*|\vec \theta)$ is injective, for a sufficient number of nuisance parameters $\vec \theta$.

Henceforth, we assume that there are simple relations (described in more detail below) between $\mbh$ and the mass when helium burning commences $\mhb$, and between $\mhb$ and the zero-age main sequence mass $\mzams$. With these assumptions, \Eq{chain-rule-1} can be written
\begin{eqnarray}
    \label{chain-rule-2}
    \frac{d\nbh^{\rm(1g)}}{d \mbh} = \int d \vec \theta \, \frac{dN_{\rm ZAMS}}{d \mzams}  \frac{ d \mzams(\mhb|\vec \theta)/d \mhb}{d \mbh(\mhb|\vec \theta) / d \mhb}  \cal P(\vec \theta).~~~~~
\end{eqnarray}
This is convenient because the ZAMS mass function $dN_{\rm ZAMS}/d \mzams$ is the stellar IMF, which at large masses can be approximated as a power law \citep{1955ApJ...121..161S, Chabrier:2003ki}, and $d\mbh(M_{\rm hb}|\vec \theta)/dM_{\rm hb}$, which describes the complicated physics of (post-)helium burning evolution, can be calculated using stellar structure simulations. The results shown in this work were computed using {\tt MESA} version 12778 \citep{Paxton:2010ji,Paxton:2013pj,Paxton:2015jva,Paxton:2017eie}. The remaining function $d \mzams( \mhb|\vec \theta)/d \mhb$ describes the astrophysics of main sequence exit, which over sufficiently long timescales we will approximate as a simple power law. This approximation does not capture the full complexity of mass loss on the main sequence \citep{2017A&A...603A.118R}; however, the weighted integral over $d \vec \theta$ combined with the relatively low number of events in GWTC-2 suggests that any departure from the power-law behavior that we assume will be reflected as a scatter around a central power law. Thus, the product $\frac{dN_{\rm ZAMS}}{d \mzams} \frac{d \mzams}{d \mhb} \cal P$ combines to a power law with some scatter, which we assume to be small.

The calculation of $\mbh$ as a function of $\mhb$ is described in detail in \citet{Croon:2020oga}. 
As the BH mass function in Eq.~\eqref{chain-rule-2} depends on derivatives of the resulting $\mbh(M_{\rm hb}|\vec \theta)$, we fit the {\tt MESA} results by a continuous function. The behaviour can be fitted well with a seven-parameter function: a constant, two power-laws with arbitrary coefficients to model both the BHs unaffected by pair-instability and PPISN BHs, and an exponential fall-off capturing PISN (see the left panels of Fig.~\ref{fig:mesaMF}). 
We further assume that the product of the stellar IMF and mass loss on the main sequence is approximated well by a simple power law, as discussed above. 
Remarkably, we find that the resulting $dN/d\mbh$ in the vicinity of the anticipated PPISN peak can be approximated well by a function of just three parameters:
\beq \label{nbh-1g}
\frac{d \nbh^{\rm(1g)}}{d \mbh} \! \propto \! \mbh^b \! \bL 1 \! + \! \frac{2 a^2 \mbh^{1/2}
(\mbhmg - \mbh)^{a-1}}{ \mbhmg^{a-1/2}} \! \bR 
\eeq
valid for $\mbh < \mbhmg$. The parameter $a$ in Eq.~\eqref{nbh-1g} determines the sharpness of the peak in the mass function, while the parameter $b$ determines the event rate as a function of mass. We smooth the ``turn-on'' of the mass function at low masses through multiplication by a function $S(\mbh|\mmin,\delta_m)$ that smoothly vanishes below a mass $\mmin$ with a width $\delta_m$, reflecting the inefficiency of black hole formation in low-mass stars \citep{Sukhbold2016ApJ...821...38S, Ertl2020ApJ...890...51E, Patton2020MNRAS.499.2803P}. We use 
\begin{equation} \label{eq:smooth-turn-on}
    S(x|y,z)
    =  \bL\exp \pL \frac{z}{x-y} + \frac{z}{x-y-z} \pR +1 \bR^{-1},
\end{equation}
for black holes in the range $\mmin \leq \mbh \leq \mmin+\delta_m$, and $S=0 (1)$ for $\mbh<\mmin$ ($\mbh>\mmin + \delta_m$), following the analysis in \citet{Abbott:2020gyp}.

We show some examples of {\tt MESA} output compared to this functional form in Fig.~\ref{fig:mesaMF}, including deviations from the Standard Model prediction due to varying the ${}^{12}{\rm C} (\alpha, \gamma)^{16}{\rm O}$ rate (the largest source of uncertainty in the standard calculation \citep{Farmer:2019jed, 2020ApJ...902L..36F}, see \ref{app:rateCO}) and additional losses from novel particles (labeled xBSM). The latter highlights the flexibility of the parametrization to capture scenarios in which PPISN is suppressed (the specific example shown is a hidden photon with mass $m_{A'}=0.01$ eV and whose kinetic mixing with Standard Model photons is determined by the parameter $\epsilon=3\times10^{-7}$ \citep{Croon:2020oga}). 
This model has a less pronounced peak, implying a smaller parameter $a$ in Eq.~\eqref{nbh-1g};
we show further examples of the dependence of the mass function on $a$ in App.~\ref{app:model}.
This peak reflects the fact that, due to PPISN, 
a wide range of stellar masses results in a narrow range of black hole masses. This is the primary physical effect that we wish to emphasize in this work. Because there is a stellar mass $\mhb^{(p)}$ that maximizes the black hole mass, the derivative of $\mbh$ as a function of $\mhb$ vanishes there: $d \mbh(\mhb^{(p)}) / d \mhb = 0$. Thus, the black hole mass function as defined by \Eq{chain-rule-2} will formally diverge at $\mbh(\mhb^{(p)}) \equiv \mbhmg$. 
In this way, the PPISN becomes manifest in our model as a peak in the black hole mass function. This peak will become more apparent as the catalog increases in size.
As we discuss in App.~\ref{app:model}, we truncate the mass function before this divergence to avoid numerical difficulties without changing the interpretation of any parameters.

Our model for the PPISN peak follows principles similar to those adopted by \citet{Talbot:2018cva}, but is more succinct, and has the virtue of cleanly ``factorizing'' the physics of the PPISN from the physics of the mass function turn-on and also cleanly separating the physics of first-generation and higher-generation black holes.
Our \Eq{nbh-1g} has only the single parameter $a$ describing the importance of PPISN, which describes the mass function near the maximum of the first-generation black hole population.  
This parameter $a$ may be used to quantify the number of black holes that are the result of PPISN, as well as identify a mass for which PPISN becomes important, as shown in more detail in App.~\ref{app:model}. 
The parametrization in Eq.~\eqref{nbh-1g} is sufficiently flexible to account for many different effects that might impact the onset of PPISN and thereby lead to a smooth change in the value of $\mbhmg$, such as variations in metallicity or wind-loss rate \citep{Farmer:2019jed, Vink2021MNRAS.tmp..894V}, a change in the nuclear reaction rates 
\citep{Farmer:2019jed, 2020ApJ...902L..36F, Woosley:2021xba}, or new physics \citep{Croon:2020ehi, Croon:2020oga, Sakstein:2020axg, Straight:2020zke, 2020arXiv201000254Z}. 

\begin{figure*}[t]
\begin{center}
\includegraphics[width=0.485\textwidth]{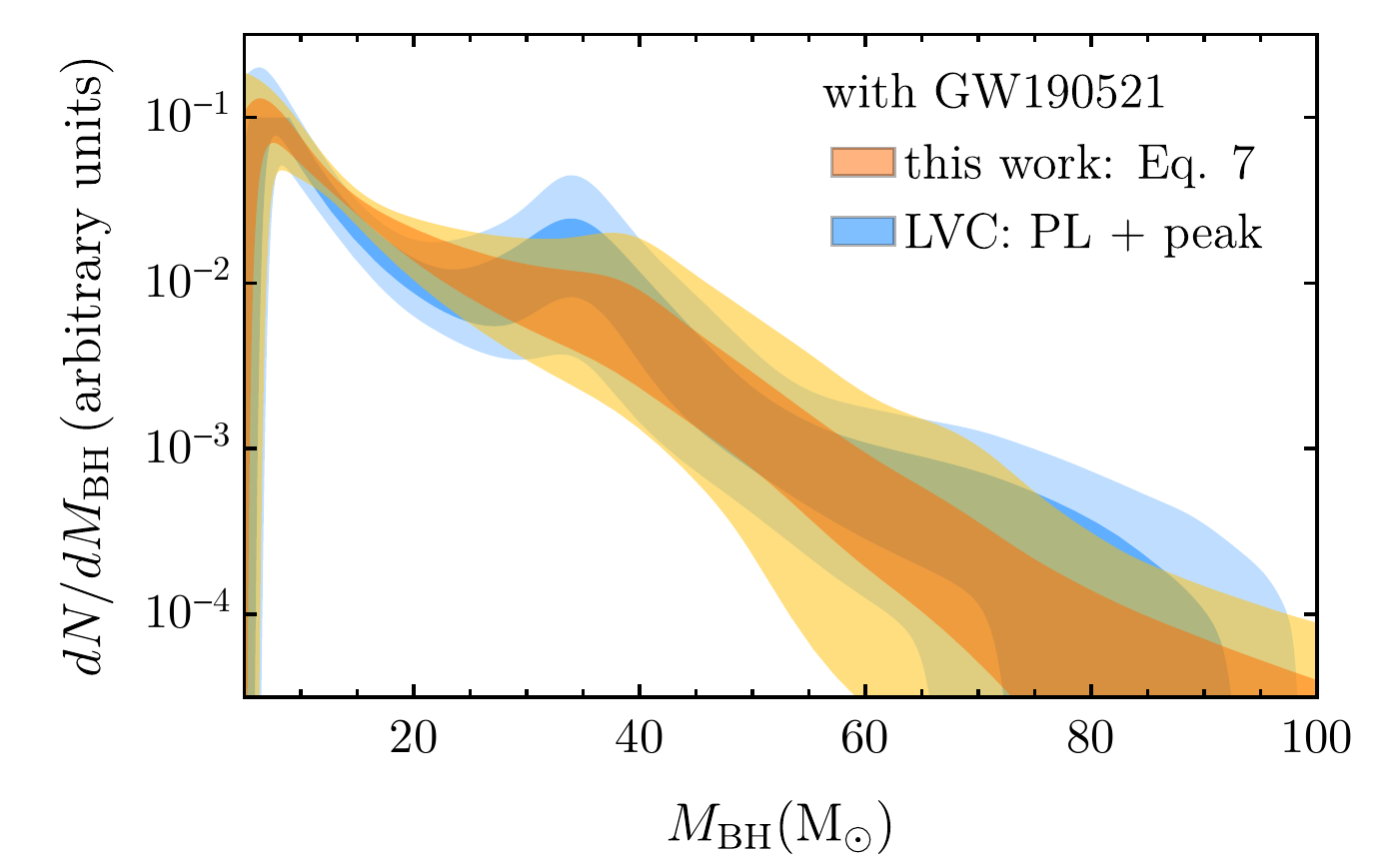}~~
\includegraphics[width=0.485\textwidth]{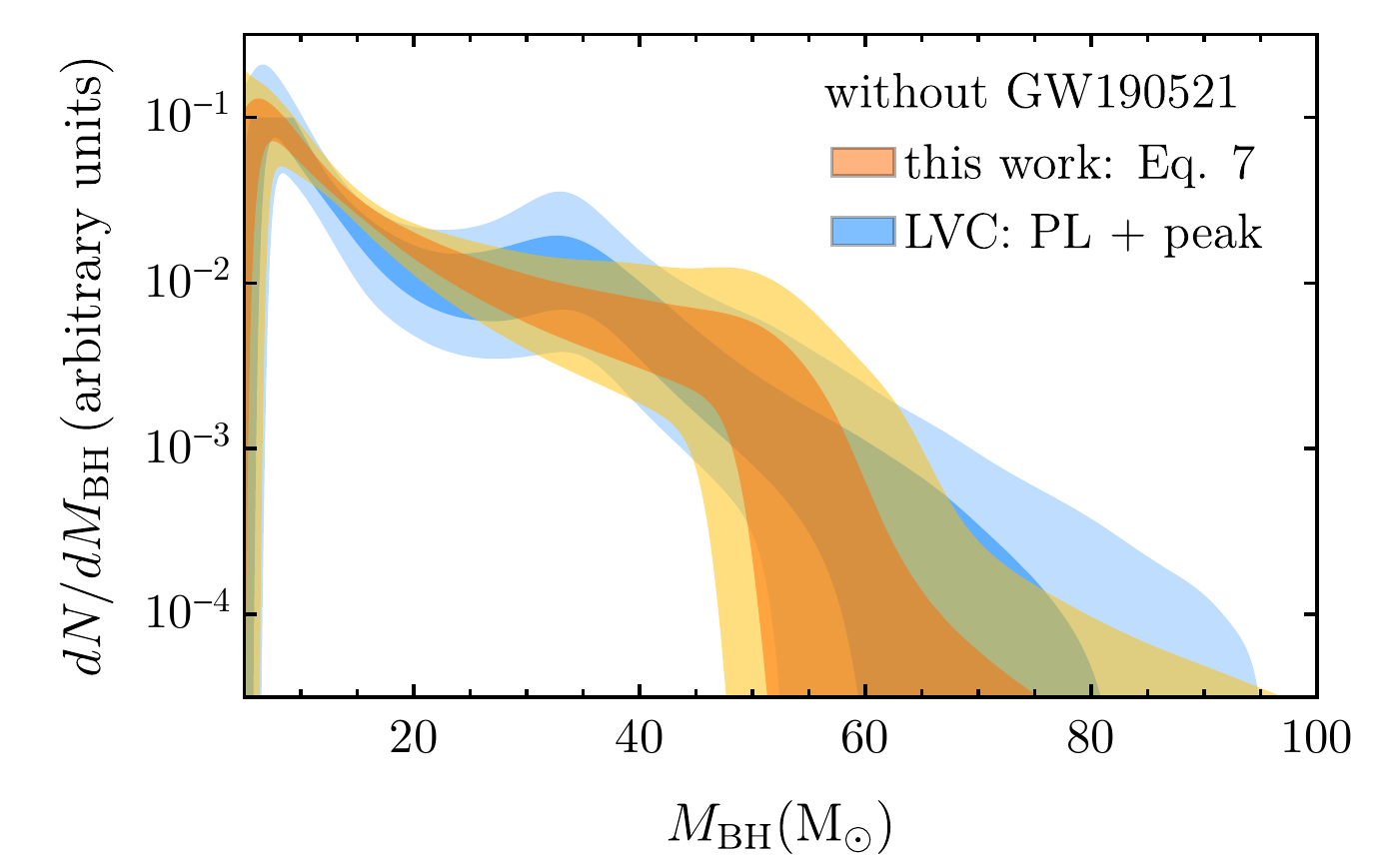}
\caption{
We compare the constraints on our mass function to constraints on the power law and peak (PL + peak) model from the LVC analysis of \citet{Abbott:2020gyp} using all BH masses inferred from GWTC-2.  We show results when the black holes associated with event GW190521 are included (left panel) and excluded (right panel).
The shaded regions are 68\% and 95\% credible intervals of many samples from the posteriors on the parameters in each fit.
}
\label{bhmf-posterior}
\end{center}
\end{figure*}

Black holes formed in prior BBH mergers \citep{Bellovary:2015ifg, Antonini:2018auk, Rodriguez:2019huv, Gerosa:2019zmo, DiCarlo:2019pmf, Yang:2019cbr, 2020ApJ...893...35D, 2020ApJ...900..177K, McKernan:2020lgr, Gonzalez:2020xah, Fragione:2020han, Weatherford:2021zdf} are not captured by the isolated black hole mass function in Eq.~\eqref{nbh-1g}. These can produce significant numbers of black holes with mass larger than $\mbhmg$ \citep{2002MNRAS.330..232C, 2017PhRvD..95l4046G, 2017ApJ...840L..24F, Rodriguez:2019huv, Zevin:2020gbd, 2020arXiv201105332K, Rodriguez:2021nwd}. 
Assuming that the rate of black hole mergers is independent of the progenitor masses, second generation black holes of mass $\mbh$ which form from the merger of two lighter first-generation black holes of mass $M_a,M_b$ will be distributed according to $ \frac{dN^{\rm(2g)}}{d \mbh} \propto \int dM_a dM_b \frac{dN^{\rm(1g)}}{dM_a} \frac{dN^{\rm(1g)}}{dM_b} \delta(\mbh - M_a-M_b)$. 
It is trivial to resolve the delta function by integrating over one of the masses, leading to
\begin{equation} \label{nbh-2g-formal}
    \frac{dN^{\rm(2g)}}{d \mbh} \propto \int dM_a \frac{dN^{\rm(1g)}}{dM_a} (M_a) \frac{dN^{\rm(1g)}}{dM_a}(\mbh -M_a) .
\end{equation}
Though the higher-generation black hole merger rate will not be completely negligible, we nevertheless emphasize that such objects are rare by assumption: not {\it every} first-generation black hole merger product will be involved in a higher-generation merger within a Hubble time. In addition, properly accounting for the contributions of these higher-generation black holes to the merger rate will entail accounting for mass- and environment-dependence related to the efficiency of formation of higher-generation binary systems \citep{2019PhRvD.100j4015C, 2020ApJ...894..133A, 2021arXiv210107699F}. For the parameters of interest in this work, we find that Eq.~\eqref{nbh-2g-formal} has a discernible two-sided peak at $\mbhmg + \mmin + \delta_m/2,$ where $\mmin$ and $\delta_m$ describe the ``turn-on'' of the black hole mass function in Eq.~\eqref{eq:smooth-turn-on}. 
We show an example of this two-sided peak in Fig.~\ref{two-sided-compare} in the appendix. We predict that such a feature will become apparent in future black hole population catalogs, but at this time we expect that this ``pollutant'' population will be subdominant to first generation black holes.
Following \citet{Abbott:2020gyp}, we introduce no further modeling to include black holes formed in BH-NS mergers at the low mass end.

Integrating Eq.~\eqref{nbh-2g-formal} is numerically intensive, since it relies on the entirety of the primary black hole mass function, and, as we discuss in more detail below, fewer than a percent of black holes are favored to originate from this population. Moreover, the width of this secondary peak is smaller than the error bars of the most massive black holes in GWTC-2. 
Thus, we adopt a simple prescription for ``pollutant'' black holes in this study:
\begin{equation} \label{nbh-2g}
    \frac{dN^{\rm(2g)}}{d \mbh} \propto  \min \bL 1,  \pL\frac{\mbh}{\mbhmg + \mmin + \delta_m/2} \pR^d \bR.
\end{equation}
This is a constant for $\mbh < \mbhmg + \mmin + \delta_m/2$ and (for $d<0$) falls monotonically at higher masses. 
An important future step will be to incorporate the stellar physics embedded in our prescription (Eq.~\eqref{nbh-2g-formal}) into the description of the pollutant population. In addition to second-generation black holes, this population will contain objects with significant post-collapse accretion \citep{vanSon:2020zbk, Belczynski:2020bca} and also black holes formed after non-isolated, pre-collapse stellar mergers \citep{DiCarlo:2019fcq, Kremer:2020wtp, Renzo:2020smh}, all of whose mass functions and contributions to the merger rate should eventually be modeled appropriately and independently. Likewise, including primordial black holes (PBHs) will require a different population model \citep{2021JCAP...03..068H, deluca2021arXiv210203809D}.

In sum, our combined mass function is 
\begin{eqnarray} \label{nbh-all}
    \frac{d \nbh}{d \mbh} &\propto& \frac{d \nbh^{\rm(1g)}}{d \mbh}  \Theta(\mbhmg-\mbh) +  \\ &~&\qquad~~ + \lambda \frac{dN^{\rm(2g)}}{d \mbh} \Theta(\mbh - \mbhmg) ,\nonumber
\end{eqnarray}
which is described by seven parameters. These parameters factorize in a physically intuitive way: $\mmin$ and $\delta_m$ describe the low-mass smoothing; $a,b,$ and $\mbhmg$ describe the first-generation black hole population; and $\lambda$ and $d$ describe the pollutant population.
We propose \Eq{nbh-all} as a physically motivated model of the black hole mass function. 
This transparently captures essential features of the astrophysical black hole population deriving from the unique physics of PPISN and PISN with a single mass scale, $M_{\rm BHMG}$.

\section{Results}
We use the BBH events detected by LVC to constrain the parameters of our population model.  We compute a model posterior following the techniques of \citet{Fishbach:2017zga}, updating the strain sensitivity \citep{2020PhRvD.102f2003B} and substituting the model of \Eq{nbh-all} for the single truncated power-law model explored there.   We note that our characterization of the LIGO selection function follows the semi-analytical treatment in \citet{Fishbach:2017zga} (also adopted in \citealt{Fishbach:2018}), rather than using the injection campaign discussed in \citet{Abbott:2020gyp}.  We do not expect this to make a large difference to our results, as evidenced by the fact that our posterior on the PL+peak model (see below) agrees quite well with that reported in \citet{Abbott:2020gyp} for the same model.  
We adopt a uniform prior on the mass ratio parameter $q \equiv m_2/m_1$, where $m_1$ and $m_2$ are the primary and secondary component masses, respectively.  While \citet{Abbott:2020gyp} adopt a power law prior on $q$ with free index $\beta_q$, this work found that the data are consistent (at roughly $1\sigma$) with $\beta_q = 0$, i.e. consistent with our assumed prior.
We do not expect the main results presented here to depend sensitively on this prior. 
However, as explored in more detail below, the inclusion or exclusion of a single event, GW190521, does strongly impact our results.

Aside from this caveat regarding GW190521, our event selection matches that of \citet{Abbott:2020gyp}.  In particular, we use all events from the O1, O2 and O3a data, with the exception of GW170817 and GW190425 (likely binary neutron star mergers), GW190814 (secondary may be a neutron star \citep{gw1908142020ApJ...896L..44A}), and GW190909\_114149, GW190719\_215514, and GW190426\_152155 (all three have false alarm rates greater than one per year). Because GW190521 is a potential outlier,
we repeat the analysis with and without GW190521.
In fact, the inference on the masses of the progenitors in GW190521 has been argued to be prior-dependent, and this event may be an intermediate-mass-ratio system that ``straddles'' the black hole mass gap \citep{Fishbach:2020qag, Nitz:2020mga}. 
If this is so, the primary progenitor object would be ``beyond'' the mass gap, while the secondary progenitor would be in the middle of our first-generation population, and thus omitting both progenitors from our analysis is similar to including the ``straddling binary''-compatible mass. As in \citet{Fishbach:2017zga}, we use the full posteriors on $m_1$ and $m_2$ reported by LIGO \citep{LSCdata} in our analysis.

We use flat priors in the ranges $20\,\msun \leq \mbhmg \leq 120\,\msun$, $0\leq a \leq 1/2$, $-4 \leq b \leq 0$, $-7 \leq \log_{10} \lambda \leq -0.3$, and $-10 \leq d \leq 0$. 
We sample the posterior and compute Bayesian evidences using a nested sampling algorithm implemented in \texttt{dynesty} \citep{Speagle:2020}.  We test our results with the affine-invariant ensemble sampler \texttt{emcee} \citep{Foreman-Mackey:2013}, which implements the proposal of \citet{Goodman:2010}.

We show constraints on the mass function of Eq.~\eqref{nbh-all} in \Fig{bhmf-posterior}. The shaded regions are 68\% and 95\% credible intervals of many samples from the posteriors of the seven parameters in our fit. These regions do not exhibit sharp peaks as the credible intervals combine many sharply peaked mass functions with different $\mbhmg$; we show some examples in App.~\ref{app:SM-vs-posterior}. We compare to the ``power law plus peak'' (PL+peak) mass function presented in \citet{Abbott:2020gyp}, for which we derive posteriors using the same priors as in \citet{Abbott:2020gyp}. The PL+peak function is a single power law slope, truncated at low and high mass, plus a Gaussian peak of variable mean, width, and height. The peak is motivated by the phenomenological form in \citet{Talbot:2018cva}, but is not constrained to lie near the end of the mass function, as we have argued is an inevitable prediction of stellar structure theory.

\begin{figure}[t]
\begin{center}
\includegraphics[width=0.485\textwidth]{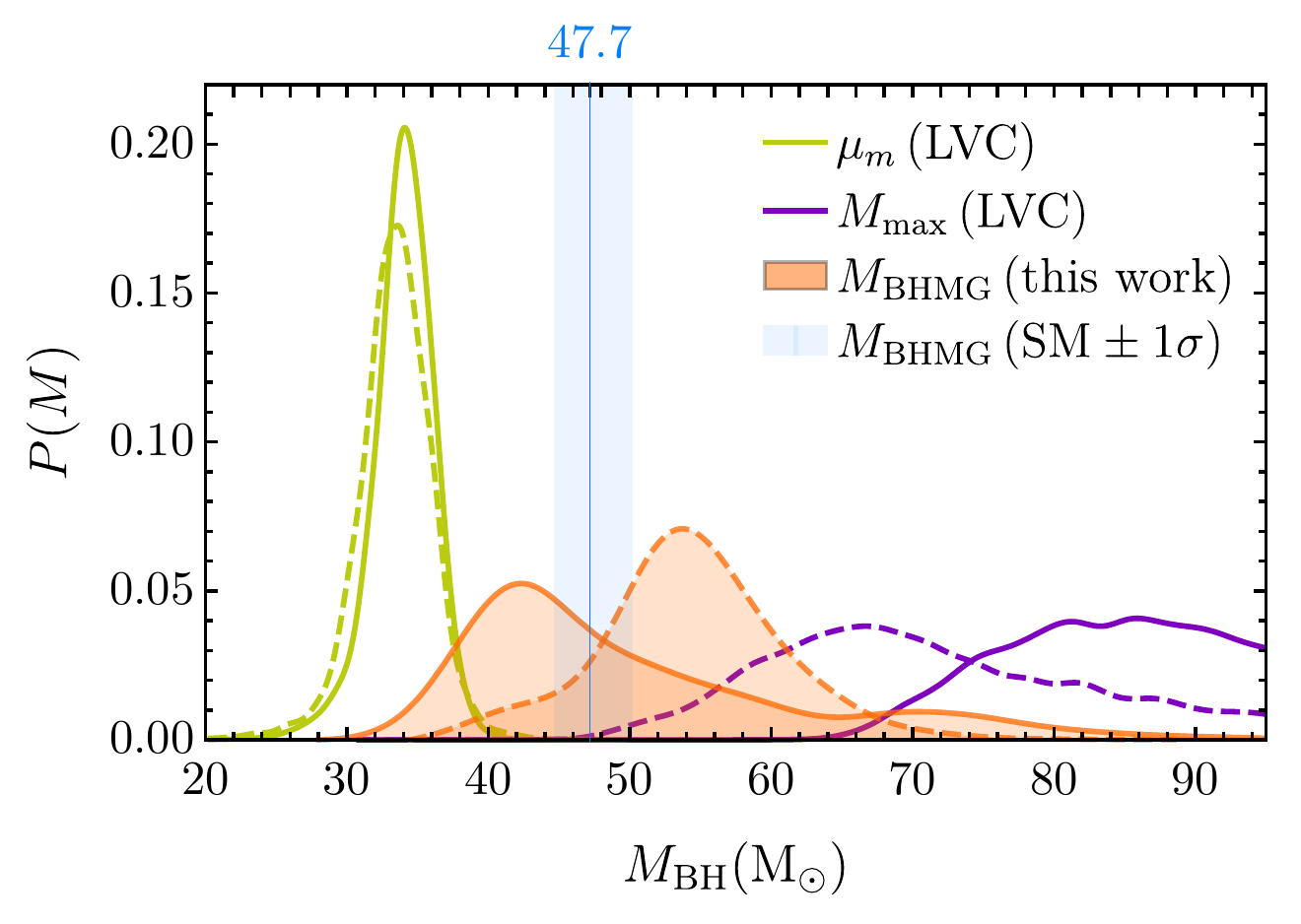}
\caption{We compare the marginalized posteriors on the various mass parameters of interest. Solid lines and histograms refer to results with GW190521; the dashed lines and the lighter shaded histograms refer to results without GW190521. 
Evidently, the inclusion of GW190521 shifts a mass parameter in both models. The blue band denotes the value of $\mbhmg$ found in the SM with $\pm 1 \sigma$ error bands of the ${}^{12}{\rm C} (\alpha, \gamma)^{16}{\rm O}$ reaction rate.
}
\label{1d-mass-posteriors}
\end{center}
\end{figure}

Excluding GW190521 from the analysis, the location of the end of the first-generation mass function remains compatible at the 1$\sigma$ level, as illustrated in Fig.~\ref{1d-mass-posteriors}. 
In Eq.~\eqref{nbh-all}, the endpoint of this mass function is interpreted in terms of a single parameter with a straightforward physical interpretation, $\mbhmg$. Without GW190521, our favored maximum first-generation black hole mass is  
$54\pm6 \,\msun$, whereas with GW190521 we conclude that 
$\mbhmg = 46^{+17}_{-6} \msun$.
The parameters describing the peak near (though not exactly at) $ \mu_m \sim 34 \rm M_\odot$ in the PL+peak model are also robust against the choice of inclusion of GW190521. A ``pile-up'' of black holes near $35 \msun$ is favored by other LVC models: the broken power law model favors a break in the power at a mass 
$m_{\rm break} = 36^{+15}_{-8} \msun $ \cite{Abbott:2020gyp}. We are not aware of any mechanism that could produce a pileup in the mass function below $40 \msun $.

If the value of $\mbhmg$ that is favored when omitting GW190521 gains in significance after future data releases, the conventional mechanism behind PPISN is favored \citep{Farmer:2019jed}, which predicts $\mbhmg = 47.7^{+3.5}_{-1}$ where the error bars given here correspond to the dominating uncertainty, from the ${}^{12}{\rm C}(\gamma,\alpha)^{16}{\rm O}$ rate \citep{Farmer:2019jed, 2020ApJ...902L..36F, deBoer2017RvMP...89c5007D,Woosley:2021xba}.
Including GW190521, the $\mbhmg$ posterior is bimodal, with a second feature near $70 \msun$. If this feature persists after future data releases, 
this will be intriguing evidence in favor of novel hypotheses that alter the location of PPISN \citep{Croon:2020ehi, Croon:2020oga, Sakstein:2020axg, Straight:2020zke}. 

We defer full corner plots on both models to App.~\ref{app:cornerplots}, where we show that all parameters aside from $\mbhmg$ and $M_{\rm max}$ are
in rough agreement regardless of the inclusion of GW190521. 
We note that the values of $\lambda$ and $\mbhmg$ are anticorrelated, which is especially apparent in the analysis with GW190521. There we see that the one-dimensional posterior on $\mbhmg$ 
is actually peaked at {\it smaller} values with the inclusion of GW190521, which has two black holes of high mass, than without it. While at first glance this may be surprising, there is in fact no contradiction, because $\lambda$ increases significantly when $\mbhmg$ is small. The second peak in the posterior for the analysis with GW190521, alluded to above, corresponds in turn to a value of $\lambda$ that is more compatible with the analysis performed {\it without} GW190521.

We note that the power-law index of the slope of the black hole mass function best-fit values and 68\% credible intervals are 
$b=-1.97\pm0.44$ ($-1.95\pm0.51$) 
excluding (including) GW190521. Each of these is compatible with measures of the stellar IMF at lower masses \citep{1955ApJ...121..161S, Chabrier:2003ki} 
which we supports our  
assumption that the overall power-law slope is inherited from the underlying stellar physics, despite our simplified treatment of mass loss on the main sequence. 

Next, the parameter $a$, which controls the number of events in the PPISN peak, is poorly constrained, and our posteriors do not favor large values of $a$.
Thus, though sharply peaked models are not substantially disfavored, neither are such models favored. This could be due to the relatively small number of high-mass events in GWTC-2, which makes it unlikely to detect the anticipated PPISN-induced ``pile-up'' of events, and may change when additional data are available.

Finally, we find that when we omit GW190521 the pollutant population power-law index $d$ is unconstrained, and the normalization of the ``pollutant'' population $\lambda$ is small, consistent with \cite{2020arXiv201105332K}. The fact that the posterior value of $\lambda$ increases and the value of $\mbhmg$ decreases when our analysis includes GW190521 suggests that this event is substantially informative for inference of these values. This is perhaps not so surprising given the as-yet relatively small number of high-mass events in GWTC-2. Future work with more data, 
and complementary modeling and simulation efforts, will enable refinements of this prescription and will presumably lead to better statistical inferences of these important physical parameters.

In order to compare the models, we compute the Bayesian evidence ${\sf E}$ for our Eq.~\eqref{nbh-all} as well as for PL+peak. The PL+peak model has the greatest evidence of the eight hypotheses compared in \citet{Abbott:2020gyp}. Without GW190521, we find that PL+peak is preferred over Eq.~\eqref{nbh-all} by $\Delta \log_{10}({\sf E}) \simeq  0.9$. This is a very mild preference, similar to the change in goodness-of-fit with the third-best models tested by \citet{Abbott:2020gyp}. 
Importantly, it performs better than the truncated model, with which this model shares a hard cut-off at the edge of $ {d \nbh^{\rm(1g)}}/{d \mbh} $.
With GW190521, our Eq.~\eqref{nbh-all} is preferred by $\Delta \log_{10}({\sf E}) \simeq 0.9$, which we take as an exciting preliminary demonstration of the validity and utility of our mass function. Given the size of the current catalog, we consider that the comparable evidence attributed to our model in Eq.~\eqref{nbh-all}, combined with its transparent physical interpretation, should encourage further use of this model for future black hole population analyses.

\section{Discussion}
We have proposed a physical model for the black hole population that facilitates transparent discovery of the signature of pulsational pair-instability supernovae in a catalog of black holes. The pair instability leads to a characteristic peak in the black hole mass function, marking the edge of the black hole mass gap, beyond which may lie a separate population. The essential physics --- that a large range of progenitor masses produce a small range of black hole masses, implying a pile up in the black hole mass catalog, beyond which there is a discontinuous and subdominant population formed in rare processes --- is transparently captured by our introduction of a single mass scale: $\mbhmg$.  The model presented here captures the importance of the pair-instability in a small number of parameters. The same qualitative features arise even if the value of $\mbhmg$ is altered by deviations of the nuclear reaction rates within their uncertainties, or if new particles drain energy from the core of the star. This means that our model provides a quantitative route to distinguish between different physical scenarios predicted by particle and nuclear physics.

The Bayesian evidence for the proposed model and the current dataset GWTC-2 is comparable to the PL+peak model, even mildly preferred when GW190521 is included in the dataset. This event remains an enigma: its inclusion shifts a mass parameter in both models (see Fig.~\ref{1d-mass-posteriors}), and the normalization between the 1g and 2g mass functions in our model. Future data sets will determine whether GW190521 was a rare 2+g event, a straddling binary, or the first indication of new physics.

This study has focused exclusively on the mass function of black holes, and we have not attempted to model other physical parameters of the events or of the different astrophysical populations of black holes. However, other parameters are inferred from the waveform of each event, and they may be used to constrain the origin and nature of the black holes in the catalog (the current analysis effectively assumes the fixed distributions, given by the parameter estimation results with flat priors from the LIGO-Virgo collaboration). In particular, the spin alignment parameter $\chi_{\rm eff}$ follows a distribution that likely depends on the dominant formation channel within a sub-population. First-generation black holes may predominantly arise in scenarios with a common-envelope history, which would lead to aligned spins \citep{Kalogera2000ApJ...541..319K, Mandel2010CQGra..27k4007M, Dominik2013ApJ...779...72D, Giacobbo2018MNRAS.474.2959G, Eldridge2017PASA...34...58E, Olejak2020ApJ...901L..39O}, whereas second-generation black holes would be formed hierarchically and thereby have a more uniform spin distribution \citep{Rodriguez2016ApJ...832L...2R, Vitale2017CQGra..34cLT01V}. One might also plausibly expect characteristic differences in the redshift \citep{2021arXiv210107699F} and eccentricity \citep{2014ApJ...784...71S, 2018PhRvD..97j3014S, 2018PhRvD..98l3005R, 2019ApJ...871...91Z} distributions between first-generation and second-generation black holes. 
These will be easy to incorporate in our model, and we look forward to future work that incorporates models for these physical effects alongside the ones that we have modelled, which we have argued are revealed by the mass function.

The future of gravitational wave observations opens up the exciting new possibility of black hole archaeology. We are hopeful that the tools presented in this paper provide useful insights into our rapidly expanding knowledge of the black hole population.

\acknowledgments
We thank Selma de Mink, R.~James deBoer, Bruce Edelman, Rob Farmer, Marco Raveri, and Mathieu Renzo for useful conversations.
We thank Christopher Berry, Reed Essick, Amanda Farah, Maya Fishbach, Mike Zevin, and especially our anonymous referee for very helpful comments on the draft.
We thank the LIGO-Virgo collaboration for the excellent science they continue to produce.
Fermilab is operated by Fermi Research Alliance, LLC under Contract No. DE-AC02-07CH11359 with the United States Department of Energy. TRIUMF receives federal funding via a contribution agreement with the National Research Council Canada. 
\clearpage
This research has made use of data, software and/or web tools obtained from the Gravitational Wave Open Science Center (\url{https://www.gw-openscience.org/}), a service of LIGO Laboratory, the LIGO Scientific Collaboration and the Virgo Collaboration. LIGO Laboratory and Advanced LIGO are funded by the United States National Science Foundation (NSF) as well as the Science and Technology Facilities Council (STFC) of the United Kingdom, the Max-Planck-Society (MPS), and the State of Niedersachsen/Germany for support of the construction of Advanced LIGO and construction and operation of the GEO600 detector. Additional support for Advanced LIGO was provided by the Australian Research Council. Virgo is funded, through the European Gravitational Observatory (EGO), by the French Centre National de Recherche Scientifique (CNRS), the Italian Istituto Nazionale di Fisica Nucleare (INFN) and the Dutch Nikhef, with contributions by institutions from Belgium, Germany, Greece, Hungary, Ireland, Japan, Monaco, Poland, Portugal, Spain.
DC and SM thank the Aspen Center for Physics, supported by NSF grant PHY-1607611, for (virtual) hospitality during the completion of this work.

\renewcommand{\eprint}[1]{\href{http://arxiv.org/abs/#1}{#1}}
\bibliographystyle{yahapj}
\bibliography{biblio}

\appendix

\section{Population model}
\label{app:model}
\subsection{Primary population} 
We model the population of primary, first generation black holes by Eq.~\eqref{nbh-1g}. This flexible parametrization is inspired by {\tt MESA} studies of PPISN, and can account for different scenarios, as demonstrated in the main text. We note that the best-fit parameters are not particularly sensitive to weight factors applied to {\tt MESA} data points. We give some further insight into this population model here, in particular into the dimensionless parameters $a$ and $b$. 

\begin{figure}[b]
\begin{center}
\includegraphics[width=0.485\textwidth]{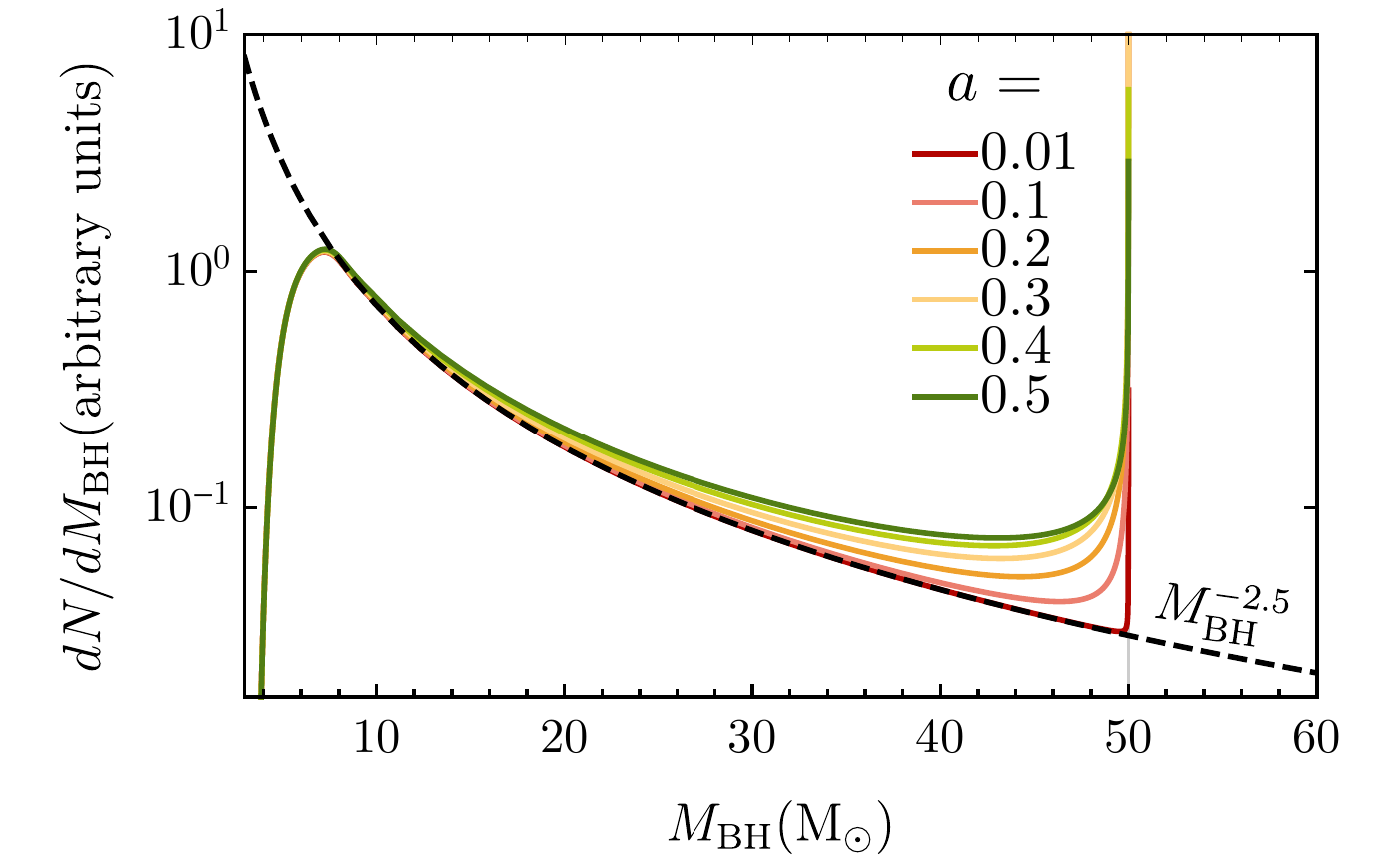}
\includegraphics[width=0.485\textwidth]{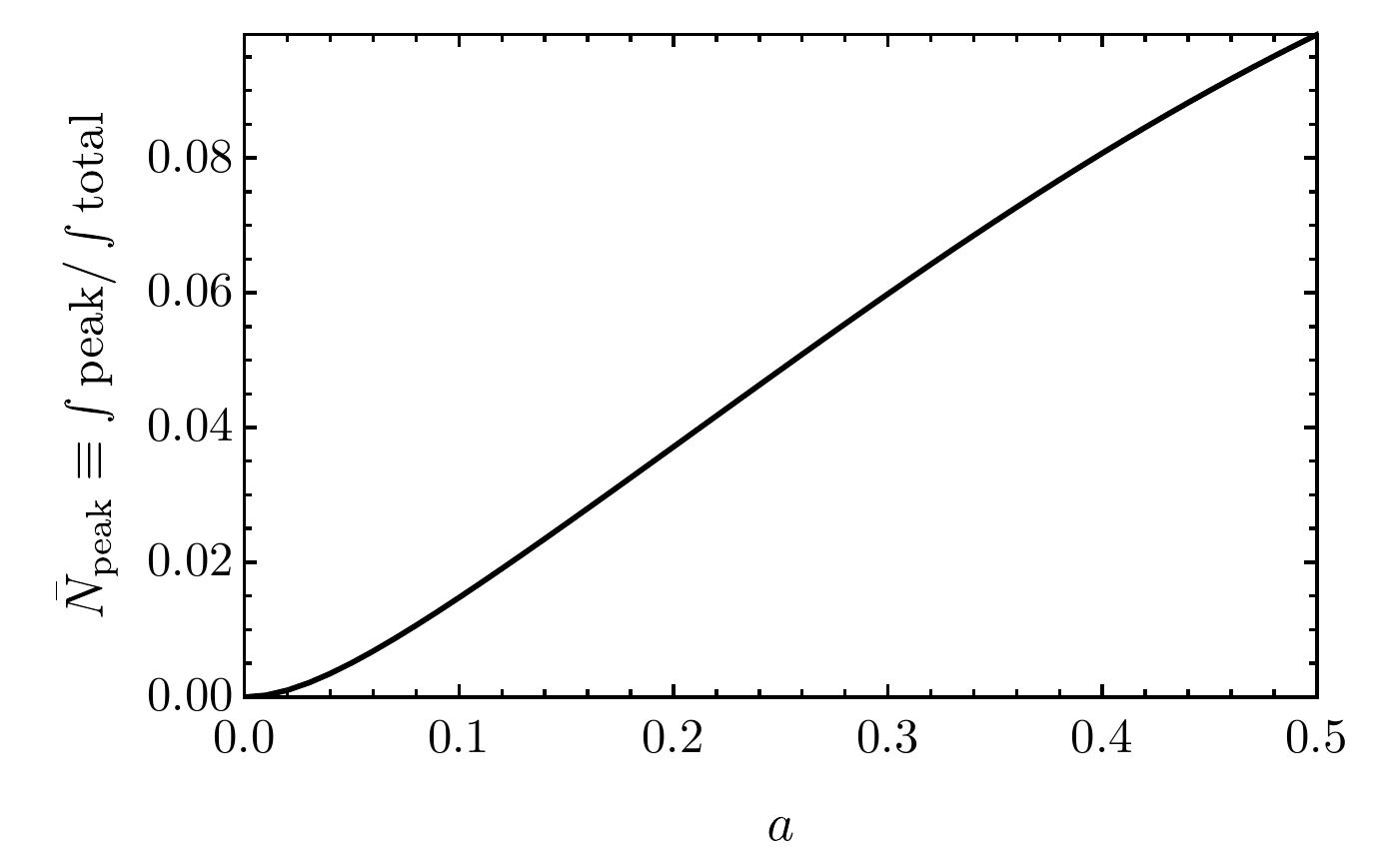}
\caption{First generation black hole mass functions parametrized by Eq.~\eqref{nbh-1g}. \textit{Left:} the effect of the parameter $a$ is demonstrated for $b=-2.5$ and $M_{\rm BHMG}=50\rm M_\odot$. \textit{Right:} for the same fiducial parameters, the fractional number of PPISN black holes is shown as a function of $a$.
}
\label{fig:primmodel}
\end{center}
\end{figure}

The spectral index $b$ describes the behaviour of the mass function at small BH masses, and is primarily informed by the stellar IMF at the beginning of helium burning. It is expected to be a negative number, informing the prior choice for this parameter. The parameter $a$ then describes the importance of the pair-instability peak. We demonstrate the effect of varying $a$ for constant $b$ in Fig.~\ref{fig:primmodel}. As the parameter $a$ varies over the prior range $[0,0.5]$, the width of the peak changes, as is seen in the left panel. In the limit $a\to 0$, the \emph{pulsation} pair-instability plays a negligible role --- stars with helium depletion masses below the mass gap lose negligible mass due to pulsations, and stars with helium depletion masses above it result in no remnant. The opposite limit $a \to 0.5$ implies pair-instability affects lighter stars, from $M\sim 30 \rm M_\odot$ for $b = -2.5$. A mild correlation in the parameters $a$ and $b$ can be expected in this limit.

As is clear from Eq.~\eqref{nbh-1g} (and as made visible in Fig.~\ref{fig:primmodel}), the distributions diverge in the limit $\mbh \to \mbhmg$. This leads to numerical instability in our sampler, so we truncate our 1g population at $(1-\ep) \mbhmg$. We set $\epsilon=0.01$ in our fiducial analysis.  Over the range of parameter values that we consider, the value of $\epsilon M_{\rm BHMG}$ is well below the uncertainties on the component black hole masses, and also well below the uncertainty on our inferred value of $M_{\rm BHMG}$. We check explicitly that this does not lead to inconsistencies when integrating the mass functions for the range of $a$ we consider. 

\subsection{Secondary Population}
In addition, we provide some further context for the parametrization of the secondary population as in Eq.~\eqref{nbh-2g}. In the absence of mass- and environment-dependent binary formation efficiency effects, the population of black holes formed in prior mergers follows a distribution described by Eq.~\eqref{nbh-2g-formal}. Ignoring these effects, an estimate of the secondary population can be made without introducing additional parameters, with the exception of a relative scaling factor. We show an example of such a computation in Fig.~\ref{two-sided-compare}. As is seen, the secondary population modeled in this way may populate the mass gap. It includes a double-sided peak near $M_{\rm BHMG}$ and falls off with a large power at larger masses. This simple result is reassuring, as it lends credibility to the expectation that hierarchical merger scenarios do not significantly erode the opportunity of measuring $\mbhmg$, given a large enough data set.

\clearpage
\section{Standard-Model Expectation versus Model Posterior}
\label{app:SM-vs-posterior}
\begin{figure}
\begin{center}
\includegraphics[width=0.485\textwidth]{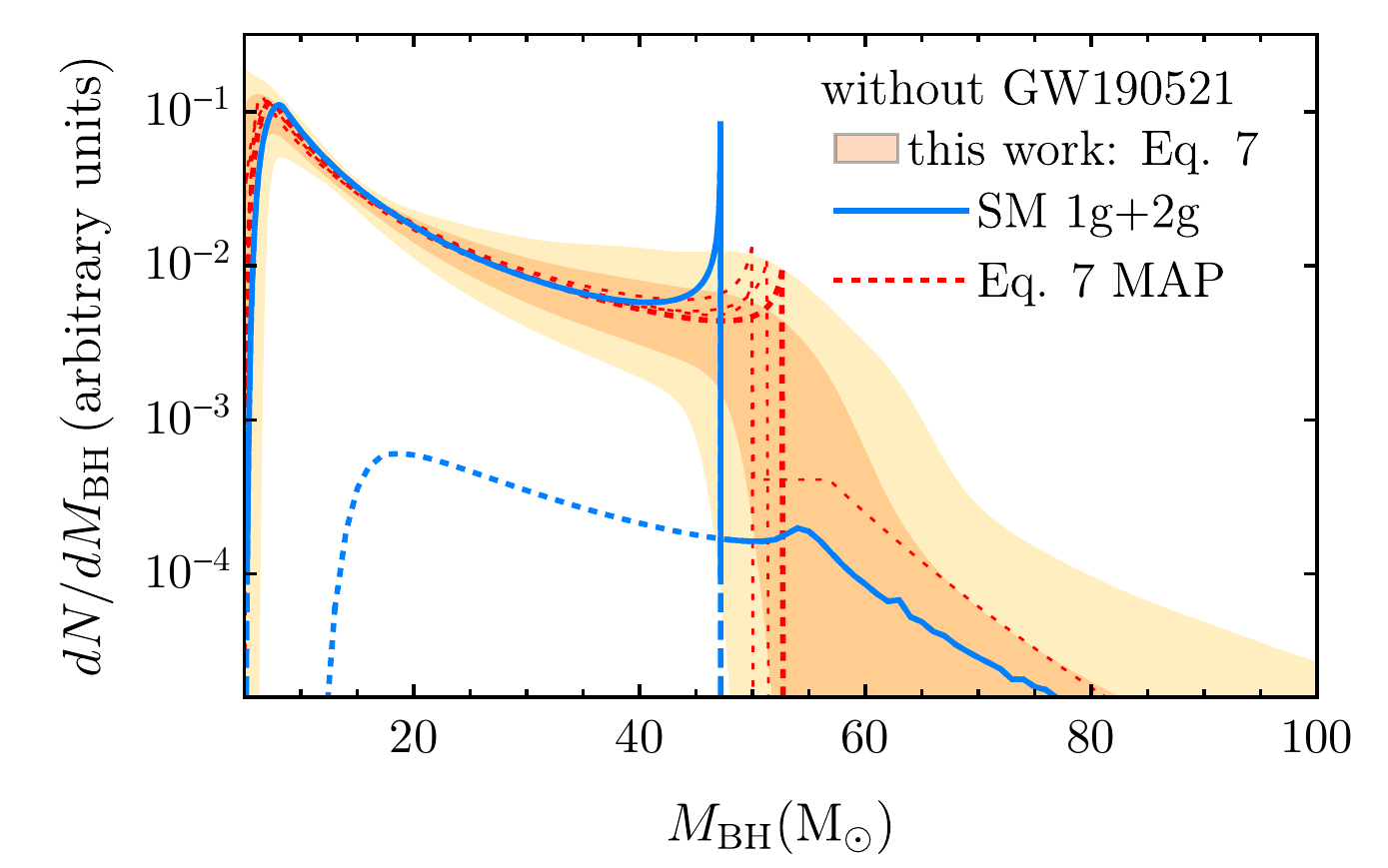}
\caption{We plot the sum (blue solid) of $dN^{\rm(1g)}/d\mbh$ (blue long dash) and $dN^{\rm(2g)}/d\mbh$ (blue short dash) as defined in Eq.~\eqref{nbh-2g-formal}. We choose the best-fit Standard Model parameters for $dN^{\rm(1g, 2g)}/d\mbh$: $a=0.39$, $b=-2.2$, $M_{\rm BHMG} = 47.2 \rm M_\odot$ as well as the best-fit values for $M_{\rm min}$ and $\delta_m$ (see table~\ref{tableparams} below), and normalize the secondary population by a factor $10^{-2}$. We compare to the 68\% and 95\% credible regions from the posterior for our analysis excluding GW190521 (orange bands), as well as individual high-probability samples from the posterior (red dashed curves).  Each of the red dashed curves has a posterior within a factor of $1/e$ of the maximum posterior (thick dashed curve).
}
\label{two-sided-compare}
\end{center}
\end{figure}

In Fig.~\ref{two-sided-compare} we show an example of a result as described in App.~\ref{app:model} for the SM mass function (see also Fig.~\ref{fig:mesaMF}). We compare this to the maximum a posteriori (MAP) result and five random samples from our posterior whose log-posterior differs from the MAP result by less than one, as well as the 68\% and 95\% credible intervals of our posterior presented in Fig.~\ref{bhmf-posterior}. Evidently, 2g mass function follows a power-law similar to the 1g function, with a small shift. It features a small peak located at $M_{\rm BHMG} + M_{\rm min} + \delta_m/2$, and falls off with a steeper power law at large mass. 
This result motivates the choice made for the more general parametrization used in the main analysis, for the specific case in which the pollutant population is composed of second generation black holes. 

\section{Uncertainties Due to the ${}^{12}{\rm C}(\alpha,\gamma)^{16}{\rm O}$ Rate}
\label{app:rateCO}
The rate of the nuclear reaction ${}^{12}{\rm C} (\alpha, \gamma)^{16}{\rm O}$ is the largest source of uncertainty for the value of $\mbhmg$ in the standard analysis \citep{deBoer2017RvMP...89c5007D, Farmer:2019jed, 2020ApJ...902L..36F, Woosley:2021xba}. The state of knowledge of this reaction is progressing rapidly in the nuclear physics community \citep{convodeBoer, Shen:2020twz}. We have implemented the $S$ factor for this reaction incorporating the latest data from \cite{deBoer2017RvMP...89c5007D} (tabulated in the reproduction package of \cite{2020ApJ...902L..36F}) in a suite of {\tt MESA} simulations. We explored the impact of one-$\sigma$ deviations from the central value on the final value of $\mbhmg$, obtaining $\mbhmg = 48^{+3.5}_{-1} \msun$. We find a central value of $\mbhmg$ that is the same as the one we find using the central value from the {\tt STARLIB} reaction library \citep{Sallaska:2013xqa}, but the extracted one-$\sigma$ error bars on $\mbhmg$ using the rates of \citet{deBoer2017RvMP...89c5007D} are almost precisely half those due to the one-$\sigma$ variations in the rate from {\tt STARLIB} \citep{Farmer:2019jed, 2020ApJ...902L..36F}.

\section{Full Results}
\label{app:cornerplots}
Below we provide our full results in the form of corner plots and tables of best fits and 68\% credible intervals for all free parameters in each hypothesis.  Corner plots and summary statistics were computed using \texttt{getdist} \citep{getdist}.

\begin{figure}[h]
\begin{center}
\includegraphics[width=0.985\textwidth]{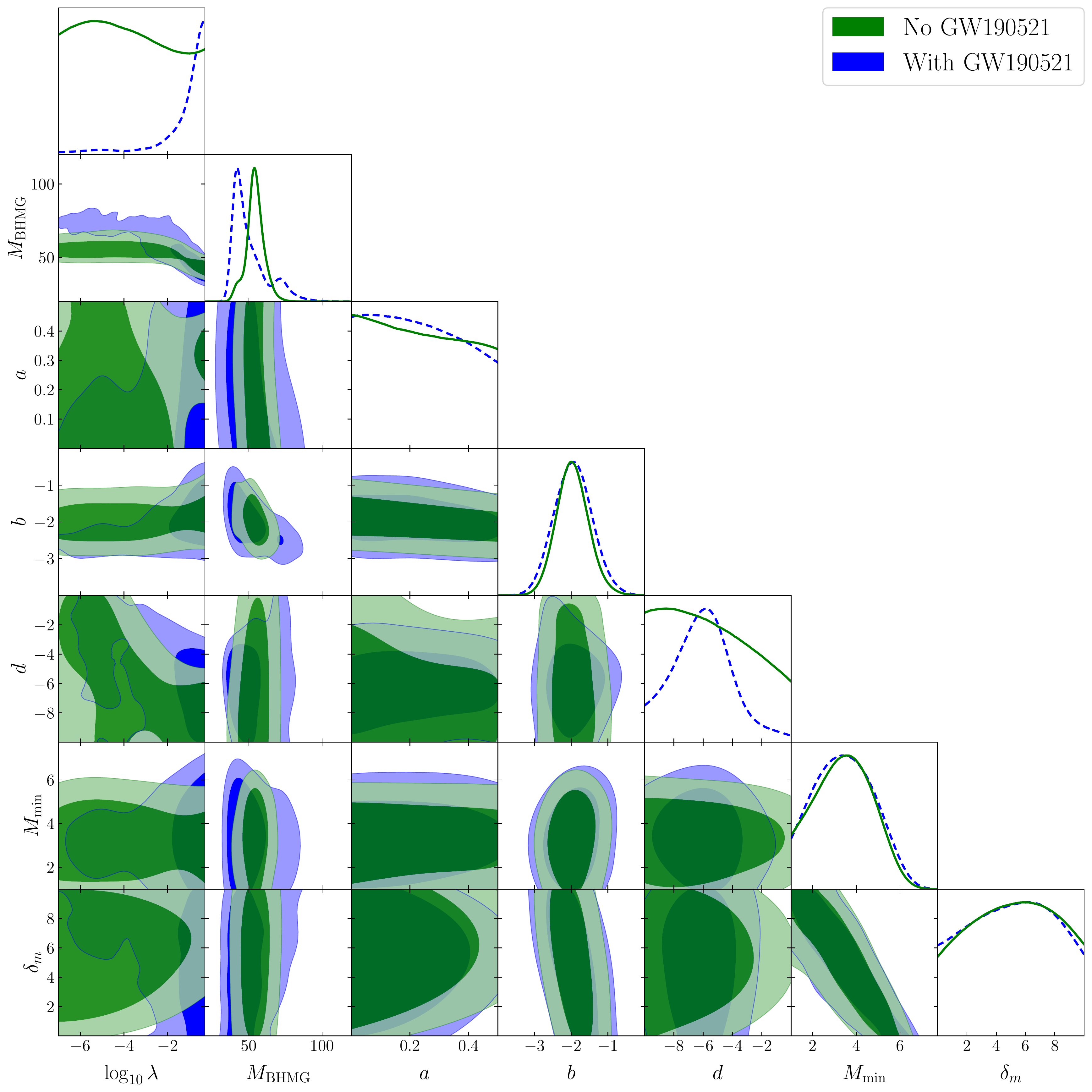}
\caption{ Posteriors on parameters from our model Eq.~\eqref{nbh-all} with and without GW190521.
}
\label{full-corner-plot-IMF}
\end{center}
\end{figure}

\begin{figure}[h]
\begin{center}
\includegraphics[width=0.985\textwidth]{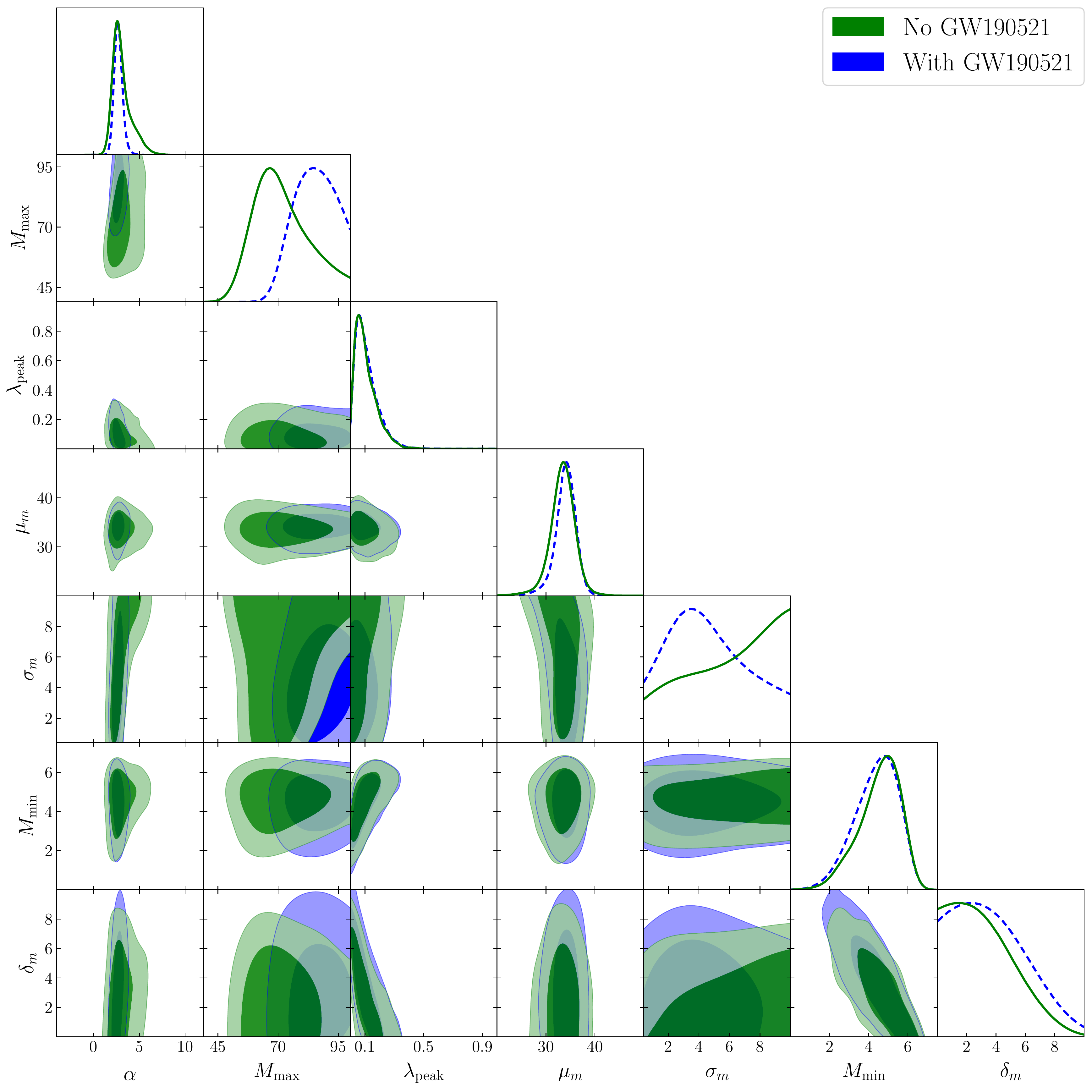}
\caption{
Posteriors on parameters from the LVC PL+peak model with and without GW190521.
}
\label{full-corner-plot-PLpeak}
\end{center}
\end{figure}

  \begin{table}[ht]
  \caption{Best fits and 68\% credible intervals for the one-dimensional marginalized posteriors on the parameters of our model in Eq.~\eqref{nbh-all} and the LVC PL+peak model. Posteriors marked by a --- indicate that these are prior-dominated.
  }
  \parbox{0.45\linewidth}{
  \centering
  \begin{tabular} { r | l | l}
   This work, Eq.~\eqref{nbh-all} &  no GW190521 & with GW190521 \\
   \hline
       $\log_{10} \lambda$ & --- & $> -1.35$ \\
       $M_{\mathrm{BHMG}}~~[\msun]$ & $54 \pm 6$ & $46^{+17}_{-6}$ \\
       $a              $ & --- & --- \\
       $b              $ & $-1.97\pm 0.44$ & $-1.95\pm 0.51$\\
       $d              $ & $< -4.10$ & $-6.0^{+1.8}_{-2.0} $   \\
       $M_{\mathrm{min}}~~[\msun]$ & $3.3^{+1.5}_{-1.7}$  & $3.3\pm 1.4$ \\
      $\delta_m~~[\msun]      $ & $5.2^{+3.0}_{-3.2}$ & $5.1^{+3.0}_{-3.2}$\\
  \end{tabular}
  }\hfill
  \parbox{0.45\linewidth}{
  \centering
  \begin{tabular} { r | c | c}
   LVC: PL+peak &  no GW190521 & with GW190521 \\
   \hline
       $\alpha         $ & $3.08^{+0.51}_{-1.2}    $& $2.72^{+0.38}_{-0.48}    $\\
       $M_{\mathrm{max}}~~[\msun]$ & $72^{+9}_{-10}          $ & $85^{+10}_{-8}           $\\
       $\lambda_{\mathrm{peak}}$ & $0.107^{+0.029}_{-0.092}  $ & $0.113^{+0.032}_{-0.094}    $\\
       $\mu_m         ~~[\msun] $ & $33.4^{+2.5}_{-2.1}      $ & $34.0^{+2.2}_{-1.7}     $\\
       $\sigma_m      ~~[\msun] $ & $> 4.49      $    & $4.7^{+1.8}_{-3.5}     $   \\
       $M_{\mathrm{min}}~~[\msun]$ & $4.56^{+1.3}_{-0.77}   $ & $4.40^{+1.3}_{-0.89}    $\\
       $\delta_m      ~~[\msun] $ & $< 4.04      $ & $< 4.75           $\\
  \end{tabular}
  \label{tableparams}
  }
  \end{table}%

\end{document}